# New Trends in Photonic Switching and Optical Network Architecture for Data Centre and Computing Systems

S. J. Ben Yoo[(1)]

[(1)] University of California, Davis, California 95616, USA, sbyoo@ucdavis.edu

**Abstract** *"AI/ML for data centres" and "data centres for AI/ML" are defining new trends in cloud computing. Disaggregated heterogeneous reconfigurable computing systems realized by photonic interconnects and photonic switching expect greatly enhanced throughput and energy-efficiency for AI/ML workloads, especially when aided by an AI/ML control plane.*

## Introduction

Global data centre IP traffic grew 11-fold over the past eight years at a Compound Annual Growth Rate (CAGR) of 25%, exceeding 20 Zettabytes per year by 2021 [1]. More recently, driven by the rapid increases in AI and machine learning related traffic, some estimates indicate that the annual data traffic will increase by over 400× over the next 10 years corresponding to CAGR of 82%. At the same time, the global energy consumption in data centers reached 200 TWh in 2020 with a CAGR of 4.4%. However, as Fig. 3 indicates [2], [3], the cost of training AI systems is doubling every 3.4 months for typical modern AI applications, now reaching 1E4 petaflops/s-days [2]-- one round of training for the biggest models at Facebook can cost "millions of dollars" in electricity consumption [3]. These trends indicate some urgent need for fundamental changes in the data centre architecture and operation since the AI/ML oriented workload increases will far outpace the technological development backed by the Moore's Law and the Dennard's Law (the Dennard's Law already became obsolete in 2008 and the Moore's Law has significantly slowed down in recent years). Some immediate observations are as follows: (1) Data movement, especially, data ingestion for training the Data Centre AI/ML system is costly; (2) Today's Data Centres are dominantly made of the von Neumann architecture which constantly requires data movements between processing units and memory; (3) While Data Centres employ accelerators, the system architecture remain mostly homogeneous; (4) Cascaded stages of electronic switches limit throughput, adds latency, and consume energy. The problem compounds as the data centre scale up and scale out. [1]

This paper discusses a number of possible strategic solutions by introducing: (1) heterogeneous reconfigurable computing in data

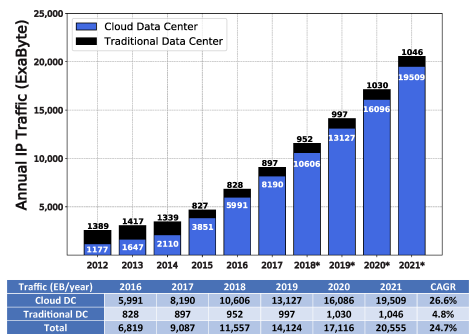

**Fig. 1.** Estimated datacenter IP traffic from 2012 to 2021 broken down by data center type. The total IP traffic is estimated to be beyond 20 exabytes per year in 2021. (Courtesy of CISCO ).

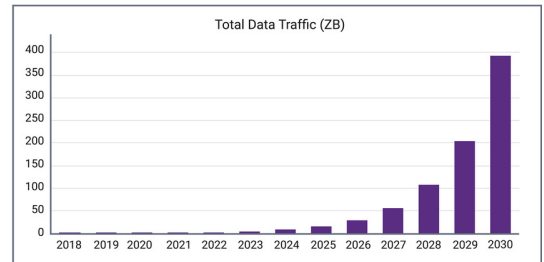

**Fig. 2.** Total Data Traffic Forecast through 2030. Source: "Impact of AI on Electronics and Semiconductor Industries", IBS, April 2020.

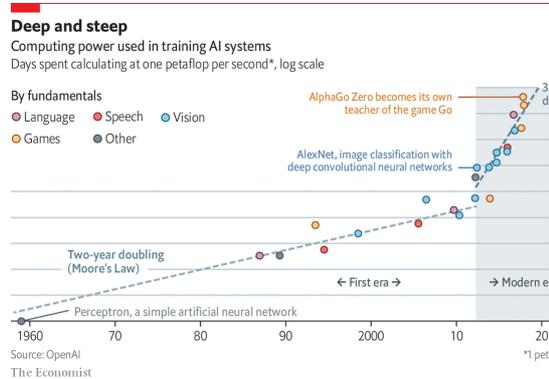

**Fig. 3.** Computing Power used in training AI Systems [Source: OpenAI.com [2], adapted in the Economist [3].]

centres enabled by photonic switching; (2) split-computing [4,5] that combines edge computing [6] and cloud computing; (3) neuromorphic computing and computing in memory (CIM); (4) a new operating system with AI/ML resource management. $3M for training transformer natural language processing [7]

**Flat, Heterogeneous, Reconfigurable Disaggregated Computing enabled by Photonic Switching and Interconnects**

It is well known that application-specific computing systems optimally designed and configured for a given workload offer much higher (often 50 or more [8,9]) energy-efficiency and throughput compared to general-purpose computing systems. Heterogeneous computing systems (HCS) exploiting energy and performance benefits of combining different domain-specific processor architectures (e.g. CPUs, GPUs, ASICs) have emerged in modern computing systems of different scales - ranging from Systems-on-chip to data center and HPC systems. Increasing numbers of processing and memory resources have led to HCSs of increasing scale, with significant computing power potential. The memory subsystem in HCS is also expected to employ a heterogeneous design with different memory technologies (i.e. volatile and non-volatile memories) used across different nodes. Furthermore, accelerators equipped with neuromorphic processors will be part of HCSs. On the other hand, application domains like graph analytics and machine learning now process terabyte data sets which are growing to petabytes. Therefore, the performance gains of running these workloads in HCSs highly depend on the scaling of the memory architecture. In an HCS, some accelerators may have embedded memory dies and some others may rely on the memory connected to the system bus. This non-uniform memory architecture (NUMA) significantly increases the complexity of programming such heterogeneous systems. Ideally, we want any compute node to be able to communicate to any memory node with sufficient bandwidth without introducing complex programming models. We want to allow programmers to reference any data from any device. Thus, it is desirable that the HCS should reconfigure its processor (i.e. CPUs, GPUs, ASICs) to memory (volatile and non-volatile memories) interconnection, following the change in the dataflow to optimize its throughput and energy efficiency across their disaggregated system. Photonic interconnects will not only offer an opportunity for disaggregated computing with extremely high energy efficiency independently of distance, but also offers an opportunity for bandwidth reconfigurable heterogeneous computing. This is true at the board-level, rack-level, and warehouse-scale computing systems. Today's data centre network architectures heavily rely on cascaded stages of many power-hungry electronic packet switches interconnected across the data centre network in fixed hierarchical communication topologies such as Fat-Tree within the data center for homogeneous racks and boards (see **Error! Reference source not found.**(a)) [10]. Due to the limited radix and bandwidth of the electronic switches, warehouse-scale data centers involve a large number of cascaded electronic switches where high energy consumption and latency compound due to repeated '*store-and-forward*' electronic processes. These architectures are also designed with a fixed topology at fixed data rates. On the other hand, as **Error! Reference source not found.**(b) illustrates, employing a passive optical fabric or a reconfigurable optical switch fabric with distributed electronic switches (e.g. ToR) could greatly improve (a) scalability of the capacity and the number of compute nodes (or racks with ToRs), (b) energy-efficiency of the network, (c) modular upgradeability, and (d) cost savings by eliminating many large and power-hungry core electronic switches at the core while keeping the smaller and disaggregated electronic switches (e.g. ToR) at the edge nodes. This transformation not only flatten the interconnect topology of the data center networks with a reduced number of hierarchies, but it also brings the possibility of optical re-configurability enhanced by wavelength division multiplexing (WDM) and silicon photonics.

**Self-Supervised vs. Reinforcement or Supervised, or Unsupervised Learning**

As Fig. 3 illustrates, AI/ML applications are now driving the cost of the energy cost of the data systems, and most of this energy cost rises from the huge amount of data ingestion required for training today's data systems based on reinforcement, supervised, or unsupervised learning mechanisms. While the AI/ML based control plane for photonic reconfigurable data systems showed promising benefits, the exponentially increasing energy requirements for training are unsustainable. On the other hand, biological systems do not require such a large amount of training data because of its self-learning (e.g. preditive-error learning [11,12]). Hence, self-supervised approach to photonic switching would be attractive [13]

**Cloud-Computing, Edge-Computing, and Split-Computing**

While cloud computing with thin client interfaces allows clients to easily access the computing resources on mobile platforms such as cell phones, it requires all data to be ingested into the centralized cloud system and casts limitations in throughput, latency, and security guarantees. For instance, future autonomous vehicles, robots, and healthcare systems may acquire 10 Tb/s data at the client machines and may need to make immediate decisions without delays or losses due to connections to the cloud system. Today's autonomous vehicles already have sophisticated GPU-based multi-sensory computing systems on board, and future systems expect to exploit even stronger computing capabilities at the edge client machines. Cloud computing will still be valuable in aggregating data and offering more global inferences based on the aggregated data. Hence, split-computing will be attractive in many of such applications, and the balance of edge vs cloud computing will depend on the use cases of various applications.

**Federated vs. Brokered Learning Systems**
Consequently, the autonomous systems at the edge can be federated into a network of semi-autonomous learning systems [14]. While federated systems offer more autonomy to the edge systems compared to the traditional hierarchical systems, the edge systems become semi-autonomous since the 'federation' assumes compromises in policies related to security and privacy, etc. Market-driven brokered system architecture [15–19] offers market-driven freedoms and dynamic choices of brokers (can be multiple brokers) with opportunities to customizing the service-layer-agreements with individual brokers.

**Intra-agent and Inter-agent Transfer Learning**
In such a federated or brokered multi-agent edge systems, the capability to transfer learn between similar but differing tasks within each agent or between multiple agents on similar tasks greatly reduces the time and energy it takes to train the agents [20–22] We expect that transfer learning process to be an important part of self-supervised learning if future computing systems [23].

**5G, 6G, and Elastic RF-Optical Networking**
5G and 6G networking with extremely high bandwidth and low latency enables unprecedented opportunities for distributed computing and edge computing in cooperation with the cloud. RF-photonic technologies are necessary to support such high-bandwidth, and must consider agility in the spectral (both RF and optical spectrum), temporal, and spatial domains in order to optimize the throughput, latency, and energy-efficiency. Similar flexible networking in the optical domain called elastic optical networking (EON) [24] has shown to optimize the network resources to achieve ~35% higher throughput or network resource savings than rigid WDM networks. Combining the RF-Photonic technologies with EON offered the desired temporal, spatial, and spectral agility in mobile RF networking (ERON) [25].
We expect to see a new generation of 5G, 6G networking supported by the ERON technologies.

**The Silicon Photonics in future computing systems**
Today's data centres and computing systems are rapidly employing silicon photonic transceivers due to their small form factors, energy-efficiency, manufacturability, and compatibility with the silicon CMOS ecosystem. Expansion of the current repertoire of silicon photonics to include 5G, 6G RF-Photonic silicon photonics, reconfigurable silicon photonic switches, and coherent optical silicon photonic systems will be natural next steps for future computing. Furthermore, integration of heterogeneous computing modules (GPU, CPU, TPU, DRAMs) with silicon photonics is already in progress in the form of heterogeneous integration in 2D, 2.5D, and 3D, and monolithic integration of such electronics and photonics is also in active progress. In the future, such integration may also include neuromorphic computing or quantum computing accelerators on silicon photonics.

**Summary**
Future computing and data system expects to benefit more from photonic-electronic reconfigurability of heterogeneous resources that are in the network of distributed computing systems with edge computing or split-computing. Self-learning and brokered learning with multi-agent transfer-learning can help reduce the burden of data ingestion currently doubling in volume every 3.3 months. Silicon photonics is expected to be ubiquitously integrated in mobile and fixed computing systems at the edge and at the core computing systems.

**Acknowledgements**
The author is grateful for contributions from many collaborators around the world.

**References**
1. Cisco, "Global data center IP traffic from 2012 to 2021, by data center type (in exabytes per year)," https://www.statista.com/statistics/227268/global-data-center-ip-traffic-growth-by-data-center-type/.

2. D. Amodei, D. Hernandez, G. Sastry, J. Clark, G. Brockman, and I. Sutskever, "AI and Compute," .
3. The ECONOMIST, "The cost of training machines is becoming a problem," The ECONOMIST, Technology Quarterly, Computing hardware (2020).
4. X. Chen and M. G. Xie, "A split-and-conquer approach for analysis of extraordinarily large data," Stat Sin **24**, (2014).
5. B. S. Rawal, R. K. Karne, and Q. Duan, "Split-System: The New Frontier of Cloud Computing," in *Proceedings - 2nd IEEE International Conference on Cyber Security and Cloud Computing, CSCloud 2015 - IEEE International Symposium of Smart Cloud, IEEE SSC 2015* (2016).
6. N. Benzaoui, "Beyond Edge Cloud: Distributed Edge Computing," in *Optical Fiber Communication Conference (OFC) 2020*, OSA Technical Digest (Optical Society of America, 2020), p. W1F.6.
7. E. Strubell, A. Ganesh, and A. McCallum, "Energy and policy considerations for modern deep learning research," in *AAAI 2020 - 34th AAAI Conference on Artificial Intelligence* (2020).
8. W. J. Dally, J. Balfour, D. Black-Shaffer, J. Chen, R. C. Harting, V. Parikh, J. Park, and D. Sheffield, "Efficient embedded computing," Computer (Long Beach Calif) **41**, 27–32 (2008).
9. A. Abnous and J. Rabaey, "Ultra-low-power domain-specific multimedia processors," in *VLSI Signal Processing, IX* (1996), pp. 461–470.
10. S. J. B. Yoo, "Prospects and Challenges of Photonic Switching in Data Centers and Computing Systems," Journal of Lightwave Technology 1–1 (2021).
11. R. C. O'Reilly, "Biologically Plausible Error-Driven Learning Using Local Activation Differences: The Generalized Recirculation Algorithm," Neural Comput **8**, 895–938 (1996).
12. R. C. O'Reilly, J. L. Russin, M. Zolfaghar, and J. Rohrlich, "Deep Predictive Learning in Neocortex and Pulvinar," J Cogn Neurosci **33**, 1158–1196 (2021).
13. C.-Y. Liu, X. Chen, Z. Li, R. Proietti, and S. J. ben Yoo, "SL-Hyper-FleX: a cognitive and flexible-bandwidth optical datacom network by self-supervised learning [Invited]," Journal of Optical Communications and Networking **14**, A113–A121 (2022).
14. H. Brendan McMahan, E. Moore, D. Ramage, S. Hampson, and B. Agüera y Arcas, "Communication-efficient learning of deep networks from decentralized data," in *Proceedings of the 20th International Conference on Artificial Intelligence and Statistics, AISTATS 2017* (2017).
15. L. Liu, Z. Zhu, X. Wang, G. Song, C. Chen, X. Chen, S. Ma, X. Feng, R. Proietti, and S. J. B. Yoo, "Field Trial of Broker-based Multi-domain Software-Defined Heterogeneous Wireline-Wireless-Optical Networks," in *Optical Fiber Communication Conference* (Optical Society of America, 2015), p. Th3J.5.
16. L. Sun, X. Chen, S. Zhu, Z. Zhu, A. Castro, and S. J. B. Yoo, "Broker-based Cooperative Game in Multi-Domain SD-EONs: Nash Bargaining for Agreement on Market-Share Partition," in *ECOC 2016; 42nd European Conference on Optical Communication* (2016), pp. 1–3.
17. L. Sun, S. Zhu, X. Chen, Z. Zhu, A. Castro, and S. J. B. Yoo, "Broker-based multi-task gaming to facilitate profit-driven network orchestration in multi-domain SD-EONs," in *2016 Optical Fiber Communications Conference and Exhibition (OFC)* (2016), pp. 1–3.
18. D. Marconett and S. J. B. Yoo, "FlowBroker: A Software-Defined Network Controller Architecture for Multi-Domain Brokering and Reputation," Journal of Network and Systems Management 1–32 (2014).
19. S. J. B. Yoo, "Multi-domain Cognitive Optical Software Defined Networks with Market-Driven Brokers," in *European Conference and Exhibition on Optical Communication (ECOC)* (2014), p. We.2.6.3.
20. C.-Y. Liu, X. Chen, R. Proietti, and S. J. ben Yoo, "Evol-TL: Evolutionary Transfer Learning for QoT Estimation in Multi-Domain Networks," in *Optical Fiber Communication Conference (OFC) 2020*, OSA Technical Digest (Optical Society of America, 2020), p. Th3D.1.
21. X. Chen, R. Proietti, C.-Y. Liu, Z. Zhu, and S. J. ben Yoo, "Exploiting Multi-Task Learning to Achieve Effective Transfer Deep Reinforcement Learning in Elastic Optical Networks," in *Optical Fiber Communication Conference (OFC) 2020*, OSA Technical Digest (Optical Society of America, 2020), p. M1B.3.
22. S. J. Pan and Q. Yang, "A survey on transfer learning," IEEE Trans Knowl Data Eng (2010).
23. R. Raina, A. Battle, H. Lee, B. Packer, and A. Y. Ng, "Self-taught learning: Transfer learning from unlabeled data," in *ACM International Conference Proceeding Series* (2007).
24. O. Gerstel, M. Jinno, A. Lord, and S. J. B. Yoo, "Elastic optical networking: a new dawn for the optical layer?," Communications Magazine, IEEE **50**, s12–s20 (2012).
25. H. Lu, Y. Zhang, Y. Ling, G. Liu, R. Proietti, and S. J. B. Yoo, "Experimental Demonstration of mmWave Multi-Beam Forming by SiN Photonic Integrated Circuits for Elastic RF-Optical Networking," in *2019 Optical Fiber Communications Conference and Exhibition (OFC)* (2019), pp. 1–3.